\documentstyle[preprint,prb,aps]{revtex}
\begin{document}

\draft

\title{Spin-density and charge-density excitations in quantum wires}

\author{ 
  Arne~Brataas$^1$ and 
  A.~G.~Mal'shukov$^2$ and
  Christoph~Steinebach$^3$ and 
  Vidar~Gudmundsson$^4$ and
  K.~A.~Chao$^1$ }

\address{
  $^1$ Department of Physics, 
  Norwegian University of Science and Technology, \\
  N-7034 Trondheim, Norway.
  }

\address{
  $^2$ Institute of Spectroscopy, 
  Russian Academy of Sciences, \\
  142092 Troitsk,
  Moscow Region, Russia.
  }

\address{
  $^3$ Institut f{\"u}r Angewandte Physik, 
  Jungiusstra{\ss}e 11,\\
  D-20355 Hamburg, 
  Federal Republic of Germany.
}
\address{
  $^4$ Science Institute, 
  University of Iceland, 
  Dunhaga 3, 
  IS-107 Reykjavik, Iceland.
  }

\date{\today}
\maketitle

\begin{abstract}
We study an interacting electron gas in a quantum wire within the
Hartree-Fock random phase approximation.  Vertex corrections to the
electron spin polarizability due to the electronic exchange
interaction are important giving rise to spin-density excitations
(SDE) with large oscillator strength shifted to lower energies with
respect to single-particle states.  The energy of intersubband SDE
oscillates with the number of subbands occupied and has a minimum when
a subband energy is close to the chemical potential.  Intrasubband SDE
have a linear dispersion at small wave-vectors.  The corresponding
sound velocity is reduced with respect to the Fermi velocity due to
exchange interaction within the occupied subbands and exchange
screening caused by virtual transitions to upper subbands.  For
intersubband and intrasubband charge-density excitations (CDE) vertex
corrections are of less importance.  For only a single subband
occupied the screening of CDE and SDE in the Tomonaga-Luttinger model
due to virtual transitions to upper subband is studied, where the
virtual transitions are treated within the Hartree-Fock approximation.
The calculations are in good qualitative agreement with experiments.
\end{abstract}



\section{Introduction}

In recent years progress has been made in preparation and
spectroscopic measurements of semiconductor nanostructures where
electrons are free to move in only one direction.  The parent system
in fabrication of such structures is a doped two-dimensional quantum
well with large ($\sim$$20\:$meV) separation between subbands, so that
electrons reside in the lower subbands.  By applying a gate voltage it
is possible to further confine the two-dimensional electron gas into a
one-dimensional wire.  The confinement gives rise to one-dimensional
subbands in the single-particle electron energy spectrum with a
subband separation of several meV.  This is a very clean system to
study since the mean free path and inelastic scattering lengths may be
longer than the length of the wire which is typically of the order of
some $\mu$m.  As a result interesting quantum effects may appear.  At
one-dimensional electron density about $10^6\:$cm$^{-1}$ more than one
subband can be occupied.

In Raman spectra\cite{Egeler90:1804,Goni91:3298,Schmeller94:14778}
collective spin and charge density excitations of GaAs quantum wires
were observed.  The dispersion of low-frequency intrasubband SDE and
CDE (plasmons) is clearly seen in angular resolved
spectra.\cite{Goni91:3298,Schmeller94:14778} The wave-vector
dependence of the spin-wave energy was found to be linear.  The
corresponding sound velocity is an important parameter which depends
on the strength of the exchange interaction between electrons.
Theoretical calculations based on the Hartree random-phase
approximation\cite{Li89:5860,Yu90:1496,Li91:11768,Wendler94:13607,Reboredo94:15174,Hwang94:17267}
have shown good qualitative and quantitative agreement with
experiments for the intrasubband and intersubband charge-density
excitations.

The role of the exchange interaction is crucial for the collective
intrasubband and intersubband SDE.  The direct long-range Coulomb
interaction leads to depolarization shifts of single-particle
excitations (SPE) and to the appearance of collective plasma modes.
Similarly the exchange interaction gives rise to red shifts of
spin-flip excitation energies.  As a result, collective SDE with a
large oscillator strength split off the continuum of SPE.  This has
been shown in Hartree-Fock RPA calculations in quantum well
semiconductor
structures.\cite{Ando82:3893,Katayama84:1615,Tselis84:3318,Eliasson87:5569}
In this paper we present analysis of electron excitations in a GaAs
quantum wire within the time dependent Hartree-Fock approximation.
The confinement potential was chosen to be parabolic.  We consider a
wire of a finite length much larger than the electron localization
length for electrons confined in a parabolic well.  The properties of
the system have been calculated for a wide range of electron numbers
and occupied subbands.

We are here primarily interested in the collective spin-density and
charge-density excitations that are two-particle processes.  The
Hartree-Fock approximation is well suited to describe the high energy
excitations and all virtual intersubband transitions.  The HF-RPA
equations are treated totally consistently in the numerical approach,
i.e.  by first solving for the HF single-particle states and then with
this HF basis solving the equation for the two-particle propagator
including vertex corrections.  This is important in order not to
violate conservation laws, e.g. to satisfy the generalized Kohn's
theorem for dipole intersubband charge-density excitations.  In the
dipole approximation, intersubband charge-density excitations at $q$=0
are simply given by the confinement energy since the perturbation only
couples to the center of mass motion which is decoupled from the
relative motion in a parabolic confinement.\cite{Kohn61:1242}
Therefore the intersubband charge-density excitations at $q$=0 do not
provide much information about the system.  However the intersubband
spin-density excitations depend on the relative motion of electrons
with different spin orientation and their energies are not given
simply by the confinement energy.  Indeed we found SDE energies to be
dependent on the electron density.  These energies oscillate with the
occupation of subbands and have a minimum when a new subband starts to
be filled.  For large wave vectors the Landau damping becomes
significant for the intrasubband charge-density and spin-density
excitations.

The HF-RPA, being a more or less reasonable tool for studying
intersubband excitations, is not a good approach for low energy
intrasubband SDE and CDE in quantum wires with one or only few
subbands occupied.  As known\cite{Mattis65:304} even weak interactions
between electrons in one-dimensional systems change their ground state
from the Fermi to the Luttinger liquid.  On the other hand, it was
shown\cite{Dzyaloshinskii73:411,Li92:13713} that the Hartree random
phase approximation gave the same dispersion for the low-energy
plasmons in the spinless system as the Tomonaga-Luttinger model.  One
can not expect that in a similar way HF-RPA gives correct results for
the SDE spectrum at small wave vectors when electron spins are taken
into account.  The Tomonaga-Luttinger model
(TLM)\cite{Mattis65:304,Tomonaga50:544,Luttinger63:1154,Haldane81:2585,Luther74:589,Schulz93:1864}
is more appropriate for this purpose.  However, this model is
restricted to essentially low-energy physics and cannot be applied to
quasi-one-dimensional systems where the coupling between the subbands
is siginficant.  Even if only one subband is occupied electron virtual
transitions to higher nonoccupied subbands should be taken into
account.  These transitions cause the direct and exchange intersubband
screening which renormalizes the intrasubband excitation energies.
The HF-RPA can be used for evaluation of these screening effects.  We
thus combine the two approaches and treat the low-energy intrasubband
excitations in the Tomonaga-Luttinger model, while their intersubband
screening was calculated with the use of HF-RPA.  We calculate the
direct and exchange screening as a function of the confinement energy.
It is shown that the screening effects are rather weak, stronger for
SDE than for CDE and of the order of the correction due the
backscattering caused by exchange scattering from one side of the
Fermi surface to the other for sufficiently small subband separation
($\sim$$5\:$meV).  In an experimental situation, these effects may be
just as important as the backscattering effect and they are not easy
to separate.  The screening may be varied by changing the parabolic
confinement, but keeping the electronic density constant.

The paper is organized in the following way.  Section \ref{s:model}
describes the model treating the ground state and excited states
within the Hartree-Fock random phase approximation.  Our results for
the intersubband excitations are presented in section
\ref{s:intersubband}.  The intrasubband excitations are discussed in
section \ref{s:intrasubband} including the calculation of the
intersubband screening factors within the Tomonaga-Luttinger model.
In addition the appendices show detailed evaluations of some crucial
Coulomb matrix elements and how the excitation energies in the system
can be found from a symmetric eigenvalue problem in the HF-RPA.

\section{Model}
\label{s:model}

We consider a strictly two-dimensional electron system lying in the
$x$-$y$ plane. The motion in the $z$ direction is neglected since we
assume that the actual thickness of the well is smaller than all other
relevant length scales.  We use the Hartree-Fock approximation to
reduce the many-particle Hamiltonian to a single-particle Hamiltonian
for each electron in an effective potential approximating the
electron-electron interaction.  The corresponding Hartree-Fock
single-particle propagator is shown in Fig.\ \ref{f:HFAground}.  For
the charge-density and spin-density correlation functions we use the
corresponding time-dependent approximation describing the
self-consistent linear response of the two-dimensional electron gas to
an external perturbation (Hartree-Fock random phase approximation).
We assume that the ground state is paramagnetic.  The Hartree-Fock
equation for a quasi-particle is then
\begin{eqnarray}
  [ & - & \frac{\hbar^2}{2 m^*} \nabla^2 + V_{\text{c}}(y) + 2
  \frac{e^2}{\kappa} \int d {\bbox r}^{\prime} \sum_{b} f_{b}
  \frac{\psi_{b}^*({\bbox r}^{\prime}) \psi_{b}({\bbox r}^{\prime})}
  {|{\bbox r}-{\bbox r}^{\prime}|} ] \psi_{a}({\bbox r}) \nonumber \\ & - &
  \frac{e^2}{\kappa} \int d{\bbox r}^{\prime} \sum_{b} f_{b}
  \frac{\psi_{b}^*({\bbox r}^{\prime})
    \psi_{b}({\bbox r})}{|{\bbox r}-{\bbox r}^{\prime}|}
  \psi_{a}({\bbox r}^{\prime}) = \epsilon_{a} \psi_{a}({\bbox r}) \, ,
  \label{hartree-fock}
\end{eqnarray}
where $f_{a}$ is the Fermi occupation factor for a state with orbital
quantum number $a$ (the spin is up or down) with eigenenergy
$\epsilon_{a}$ and $m^*$ is the effective mass.  The dielectric
constant of the surrounding medium is noted by $\kappa$.  We assume
that the confining potential is parabolic
\begin{equation}
  V_c(y) = \frac{1}{2} m^* \omega_0^2 y^2
\end{equation}
giving a subband spacing of $\hbar \omega_0$.  Periodic boundary
condition in the longitudinal direction of the wire gives a Bloch-type
single-particle wave function
\begin{equation}
  \Psi_{nk}(x,y) = \frac{1}{\sqrt{L_x}} e^{i k x}\psi_{nk}(y) \, ,
\end{equation}
where $n$ is the subband index and the longitudinal wave-vector is $k
$=$p$$\cdot$$2\pi/L_x$ with $p$$\in$$Z$ and $L_x$ is the length of the
wire.  For noninteracting electrons in a simple one-dimensional
parabolic potential the transverse single-particle eigenfunctions are
independent of the longitudinal wave-vector
\begin{equation}
  \phi_{n}(y) = \frac{1}{\sqrt{l_0}} \frac{1}{\sqrt{2^n n!
      \sqrt{\pi}}} H_n\left( \frac{y}{l_0}\right) \exp{\left[
      -\frac{y^2}{2l_0^2}\right] } \, ,
\label{sinwave}
\end{equation}
where the $n$th Hermite polynomial is denoted by $H_n$.  The electron
is localized in the transverse direction within the order of the
confinement length defined as $l_0$=$[\hbar /( m^* \omega_0)]^{1/2}$.
The eigenenergies corresponding to the eigenstates (\ref{sinwave}) are
\begin{equation}
  E_{nk} = \hbar \omega_0 \left(n + \frac{1}{2} \right) +
  \frac{\hbar^2 k^2}{2m^*} \, .
\end{equation}
The effective single-particle Hamiltonian corresponding to
(\ref{hartree-fock}) is diagonalized using the wave functions of the
noninteracting electrons (\ref{sinwave}) as a functional basis and the
self-consistent solutions are obtained by iteration.

For inelastic light scattering not close to resonance the Raman
intensities in polarized and depolarized scattering geometries are
proportional to the imaginary parts of the charge-density correlation
function and spin-density correlation function
respectively.\cite{Hamilton69} The charge-density correlation function
is
\begin{equation}
  \chi^{+}({\bbox q},\omega) = - \frac{i}{\hbar} \int_{0}^{\infty} dt
  e^{i \omega t} \langle |\left[\hat{\rho}({\bbox
  q},t),\hat{\rho}(-{\bbox q},0) \right] | \rangle \, ,
\label{Ccorr}
\end{equation}
and the spin-density correlation function is
\begin{equation}
  \chi^{-}({\bbox q},\omega) = - \frac{i}{\hbar} \int_{0}^{\infty} dt
  e^{i \omega t} \langle |\left[\hat{\sigma}({\bbox
  q},t),\hat{\sigma}(-{\bbox q},0) \right] | \rangle \, ,
\label{Scorr}
\end{equation}
where $\hat{\rho}({\bbox q},t)$ is the Fourier transform of the
charge-density operator and $\hat{\sigma}({\bbox q},t)$ is the Fourier
transform of the spin-density operator along the spin quantization
axis and $\langle | \ldots |\rangle$ denotes the thermodynamic
average.  In fact there are three different possible spin-density
correlation functions, however this triplet is degenerate for the
nonmagnetic system discussed here.  We denote the real part of the
correlation functions (\ref{Ccorr}) and (\ref{Scorr}) as
$R^{\pm}({\bbox q},\omega)$ and the imaginary part as
$S^{\pm}({\bbox q},\omega)$
\begin{equation}
  \chi^{\pm}({\bbox q},\omega) = -R^{\pm}({\bbox q},\omega) - i
  S^{\pm}({\bbox q},\omega) \, .
\label{reimcorr}
\end{equation}
The real and imaginary parts of retarded correlation functions are
related by the Kramers-Kronig relations which is used in the numerical
evaluation of the real part of the correlation function, see Appendix
\ref{s:diagonalization}.

If treated exactly the spectral function of the charge-density and
spin-density correlation function should satisfy the longitudinal
f-sum rule independent of the Coulomb interaction
\begin{equation}
  \int_{-\infty}^{\infty} d\omega \omega S^{\pm}({\bbox q},\omega) = \pi
  \frac{q^2}{m^*} N \, ,
\label{fsum}
\end{equation}
where $N$ is the number of electrons in the wire.  The charge-density
operator is
\begin{equation}
  \hat{\rho}({\bbox q}) = \sum_{a,b}\int d{\bbox r} e^{i {\bbox q} \cdot
    {\bbox r}} \psi_{a}^*({\bbox r}) \psi_{b}({\bbox r})
  \left(\hat{c}_{a\uparrow}^{\dag} \hat{c}_{b\uparrow}+
    \hat{c}_{a\downarrow}^{\dag} \hat{c}_{b\downarrow}\right)
\end{equation}
and the spin-density operator is
\begin{equation}
  \hat{\sigma}({\bbox q}) = \sum_{a,b} \int d{\bbox r} e^{i {\bbox q} \cdot
    {\bbox r}} \psi_{a}^*({\bbox r}) \psi_{b}({\bbox r})
  \left(\hat{c}_{a\uparrow}^{\dag} \hat{c}_{b\uparrow}-
    \hat{c}_{a\downarrow}^{\dag} \hat{c}_{b\downarrow}\right)
\end{equation}
expressed in terms of quasi-particle Hartree-Fock states where
$\hat{c}_{a\uparrow}$ destroys a particle in a state with orbital
quantum number $a$ and spin up.  Therefore the correlation functions
(\ref{Ccorr}) and (\ref{Scorr}) may be written as
\begin{equation}
  \chi^{\pm} ({\bbox q},\omega) = \sum_{abcd} \langle a | e^{i {\bbox q}
    \cdot {\bbox r}} | b \rangle \langle c | e^{-i {\bbox q} \cdot
    {\bbox r}} | d \rangle \Pi_{ab;cd}^{\pm}(\omega) \, ,
\end{equation}
where the two-particle charge-density and spin-density correlation
functions
\begin{eqnarray}
  \Pi_{ab;cd}^{\pm}(\omega) &=& \Pi_{a\uparrow b\uparrow;c\uparrow
    d\uparrow}(\omega) \pm \Pi_{a\uparrow b\uparrow;c\downarrow
    d\downarrow}(\omega) \pm \nonumber \\ & &\Pi_{a\downarrow
    b\downarrow;c\uparrow d\uparrow}(\omega) + \Pi_{a\downarrow
    b\downarrow;c\downarrow d\downarrow}(\omega)
\label{twocharge}
\end{eqnarray}
are combinations of the retarded two-particle Green's function defined
as
\begin{eqnarray}
  && \Pi_{\alpha\beta;\gamma\delta}(\omega) = \nonumber \\ && -
  \frac{i}{\hbar} \int_0^{\infty} dt e^{i \omega t}\langle | \left[
    \hat{c}_{\alpha}^{\dag}(t) \hat{c}_{\beta}(t),
    \hat{c}_{\gamma}^{\dag}(0) \hat{c}_{\delta}(0) \right] |\rangle \,
  .
\label{twoGreen}
\end{eqnarray}
Here greek indices contain orbital and spin quantum numbers.  The
two-particle Green's function (\ref{twoGreen}) are found in the
HF-RPA, which neglects correlation effects.  The fermion closed loop
and all ladder diagrams are included in the polarization operator and
the Hartree-Fock single-particle Green's function are used for the
fermion lines, see Fig.\ 
\ref{f:HFRPA}.\cite{Kallin84:5655,MacDonald85:1003,Gudmundsson95:11266}
The Hartree-Fock random phase approximation is equivalent to a time
dependent Hartree-Fock approximation.  Since the time dependent
Hartree-Fock approximation is a conserving
approximation,\cite{Baym61:287} the longitudinal f-sum-rule
(\ref{fsum}) is in general satisfied and serves as a test of the
consistency of our numerical procedure.  By using the Matsubara
technique\cite{Mahan90} (introducing the complex frequency $i \hbar
\omega_n$=$i2n\pi k_b T$ where $n$ is an integral number, $k_b$ is the
Boltzmann constant and $T$ is the temperature) to find the
two-particle Green's function at finite temperatures, the matrix
equation
\begin{eqnarray}
  & & \Pi_{\alpha\beta;\gamma\delta} (i \omega_n) =
  \Pi_{\gamma\delta}^0(i\omega_n) \delta_{\gamma,\beta}
  \delta_{\alpha,\delta} + \nonumber \\ & &
  \Pi_{\gamma\delta}^0(i\omega_n)
  \sum_{\gamma^{\prime}\delta^{\prime}} \left(
    V_{\gamma^{\prime}\delta;\gamma\delta^{\prime}} - V_{\delta
      \gamma^{\prime};\gamma\delta^{\prime}} \right)
  \Pi_{\alpha\beta;\gamma^{\prime}\delta^{\prime}}(i\omega_n) \, ,
\label{twoHF}
\end{eqnarray}
is obtained where $\Pi_{\gamma\delta}^0(i\omega_n)$ is the unperturbed
two-particle Green's function.  It is given by the single-particle
Hartree-Fock Green's functions
\begin{eqnarray}
  \Pi_{\gamma\delta}^0(i \omega_n) & = & \frac{1}{\beta \hbar} \sum_{i
    \omega_{n^{\prime}}} G_{\gamma}(i \omega_n + i
  \omega_{n^{\prime}}) G_{\delta}(i \omega_{n^{\prime}}) \nonumber \\ 
  & = & \frac{f_{\delta} - f_{\gamma}} {i \hbar \omega_n -
    (\epsilon_{\gamma}- \epsilon_{\delta})}
\end{eqnarray}
and the Coulomb matrix element is 
\begin{equation}
  V_{\alpha \gamma; \delta \beta} = \frac{e^2}{\kappa} \int d{\bbox r}
  \int d{\bbox r}^{\prime} \frac{\psi_{\alpha}^*({\bbox r})
    \psi_{\beta}({\bbox r}) \psi_{\gamma}^*({\bbox r}^{\prime})
    \psi_{\delta}({\bbox r}^{\prime})} {|{\bbox r}-{\bbox r}^{\prime}|} \, .
\end{equation}
The negative term in the summation (\ref{twoHF}) represents the vertex
corrections.  In the Hartree random phase approximation, this term is
neglected.  We now introduce the induced ``density'' matrices
$K_{cd}^{\pm}({\bbox q},\omega)$
\begin{equation}
  K_{cd}^{\pm}({\bbox q},\omega) = \sum_{ab} \langle a| e^{i {\bbox q}
    \cdot {\bbox r}} | b \rangle \Pi_{ab;cd}^{\pm}(\omega) \, ,
\end{equation}
in order that the correlation functions may be expressed as
\begin{equation}
  \chi^{\pm}({\bbox q},\omega) = \sum_{cd} K_{cd}^{\pm}({\bbox q},\omega)
  \langle c | e^{- i {\bbox q} \cdot {\bbox r}} | d\rangle \, .
\end{equation}
The charge-density induced ``density'' matrix satisfies the equation
\begin{eqnarray}
  && K_{ab}^{+}({\bbox q},\omega) = \frac{f_{b}-f_{a}}{\hbar \omega -
    (\epsilon_{a}-\epsilon_{b})} [ 2 \langle a | e^{-i {\bbox q}
    {\bbox r}} | b \rangle^* + \nonumber \\ && \sum_{cd}
  \left(2V_{cb;ad} - V_{bc;ad}\right) K_{cd}^{+}({\bbox q},\omega) ] \,
  ,
\label{indcharge}
\end{eqnarray}
and similarly for the induced spin-density matrix
\begin{eqnarray}
  & & K_{ab}^{-}({\bbox q},\omega) = \frac{f_{b}-f_{a}}{\hbar \omega -
    (\epsilon_{a}-\epsilon_{b})} [ 2 \langle a | e^{-i {\bbox q}
    {\bbox r}} | b \rangle^* - \nonumber \\ && \sum_{cd} V_{bc;ad}
  K_{cd}^{-}({\bbox q},\omega) ] \, .
\label{indspin}
\end{eqnarray}
If vertex corrections are neglected the spin-density excitation
spectra are simply given by single-particle Hartree-Fock energies.  In
HF-RPA spin-density excitations are shifted from the single-particle
Hartree-Fock energies due to the exchange interaction.  As a result
collective spin-density excitations may appear in the spectra.  The
shift from the single-particle Hartree-Fock energies is a measure of
the strength of the exchange interaction in the system.  For long
wavelength charge-density excitations the direct Coulomb interaction
dominates giving rise to plasmons, so that the exchange interaction is
not so important here.  It will however give a different coupling of
the single-particle excitations and collective excitation (Landau
damping).  The eigenequations (\ref{indcharge}) and (\ref{indspin})
have to be solved numerically.  See Appendix \ref{s:diagonalization}
for a more detailed description of the numerical approach.

\section{Intersubband excitations}
\label{s:intersubband}

In the conventional backscattering geometry for Raman
experiments\cite{Egeler90:1804,Goni91:3298,Schmeller94:14778} the
wavelength of the incident and scattered light are much longer than
the width of the wire.  The dipole approximation is therefore valid
for intersubband excitations.  Far infrared absorption measurements in
the transverse direction provide the same information as polarized
Raman scattering, which in the case of parabolic confinement is simply
that the generalized Kohn's theorem is satisfied giving a single peak in
the spectra at $\hbar \omega_0$.  Intersubband spin-density
excitations may supply interesting additional spectroscopic
information.  In the noninteracting case and within the dipole
approximation the spectral function is
\begin{eqnarray}
  S^0(q_y,\omega) & = & \left[ \delta(\hbar \omega-\hbar\omega_0) -
    \delta(\hbar \omega+ \hbar\omega_0) \right] \times \nonumber \\ &
  & \pi (\frac{\hbar^2 q_y^2}{2m^*})/ (\hbar \omega_0)
  \label{interS0}
\end{eqnarray}
and the static susceptibility is
\begin{equation}
  R^0(q_y) = (\frac{\hbar^2 q_y^2}{m^*})/ (\hbar \omega_0)^2
  \label{interR0}
\end{equation}
for both the spin-density correlation function and charge-density
correlation function.  The charge-density correlation function does
not depend on the Coulomb interaction according to the generalized
Kohn's theorem in the dipole limit.  Therefore $S(q_y,\omega)$ and
$R(q_y)$ ($q_y$$\rightarrow$0) are invariant with respect to the
electronic density for charge-density excitations.

In the calculation of the Hartree-Fock ground state, the functional
basis has been chosen large enough in order that further expansion of
it or further iteration of the Hartree-Fock equations does not result
in visual change to the single-particle energy spectra or the electron
density of the ground state.  To attain sufficient accuracy in the
calculation of the charge-density excitations and the spin-density
excitations the size of the functional basis of the excited states has
been chosen such that further refinement results in the change of the
location of the excitation peaks smaller than a typical linewidth in
experiments $\hbar \eta$$\approx$$0.1\:$meV.  For the calculations we
employ the usual GaAs parameters, $m^{*}$=$0.067m_0$, where $m_0$ is
the electron mass, and the dielectric constant $\kappa$=12.4.  The
calculations have been performed for $T$=$1.0\:$K.

We have calculated the intersubband spin-density excitations as a
function of the electron density for two different confining
potentials; $\hbar \omega_0$=$11.37\: $meV ($l_0$=$100\:$\AA) and
$\hbar \omega_0$=$7.90\:$meV ($l_0$=$120\:$\AA).  The length of the
wire in both calculations is $L_x$=$2400\:$\AA .  In all our results,
the generalized Kohn's theorem is satisfied with a high degree of
accuracy for charge-density excitations, i.e. there is only a single
peak in the spectra at $\hbar \omega_0$ having all the spectral
intensity satisfying the f-sum rule (\ref{fsum}) as given in
(\ref{interS0}).  The generalized Kohn's theorem is also satisfied
within the Hartree random phase approximation.  Therefore a
Hartree RPA and a Hartree-Fock RPA calculation will give the same
result for the long wavelength intersubband charge-density excitations
for a parabolic confined wire.

Numerical results for the spin-density excitations as a function of
the electronic density are shown in Fig.\ \ref{f:interSDE11.37} where
$\hbar \omega_0$=$11.37\: $meV and Fig.\ \ref{f:interSDE7.90} where
$\hbar \omega_0$=$7.90\: $meV.  The upper panel of Figs.\
\ref{f:interSDE11.37} and \ref{f:interSDE7.90} shows the static
spin-density susceptibility and the center graph the excitation energy
of the dominant peak in the spectra (SDE).  The lowest subfigure shows
the difference in energy between states at the zone center and the
chemical potential for different subbands ($\epsilon_{n,k=0}$-$\mu$).
States below the horizontal line are occupied.  The dominant peaks in
the spectra have intensities that are typically one order of magnitude
larger than the weaker peaks in the spectra in the density range
shown.  The intersubband SDE energy is red-shifted with respect to the
single-particle Hartree-Fock energies.  The excitation energy
oscillates with the subband filling and has local minima where a new
subbands starts to be filled.  In this case the screening is large so
that the self-consistent subband spacing is significantly smaller than
the confinement energy $\hbar \omega_0$.  In Fig.\ \ref{f:interSDE_CS}
the spin-density excitation with the largest oscillator strength has
been calculated for densities common in present experiments on
wires. {\it In this calculation only}, we use the local density
approximation (LDA) for the exchange interaction with the
parametrization\cite{Tanatar89:5005}
\begin{equation}
  V_{x}^{LDA}(y)=-2\sqrt{\frac{2\:
      n^{2D}(y)}{\pi}}\;\frac{e^2}{\kappa}\, .
\label{LDA}
\end{equation}
For high density the oscillations of the spin-density excitations are
weak.  When the 2D electronic density is sufficiently low and we are
close to a new subband starting to be filled, we see that the HF-RPA
gives overdamped modes (imaginary excitation energy, see Appendix
\ref{s:diagonalization}) for the spin-density excitation, e.g. for
$\hbar \omega_0$=$11.37\:$meV ($l_0$=$100\:$\AA ) there is an
overdamped region $n_{1D}$=$(7.1$-$8.3)$$\times$$10^5\:$cm$^{-1}$ and
for $\hbar \omega_0$=$7.90\:$meV ($l_0$=$120\:$\AA ) there are
overdamped regions
$n_{1D}$=$(4.6$-$7.5,12.1$-$14.5)$$\times$$10^5\:$cm$^{-1}$ for the
densities shown.  The presence of overdamped modes means that our
Hartree Fock ground state is not stable in these density regimes.  The
oscillations in the excitation energy are followed by peaks in the
static paramagnetic susceptibility and the static paramagnetic
susceptibility is divergent in the overdamped regions.  By comparing
the results in Fig.\ (\ref{f:interSDE11.37}) and Fig.\ 
(\ref{f:interSDE7.90}) we see that there are broader unstable regimes
for $\hbar \omega_0$=$11.37\: $meV ($l_0$=$100\: $\AA) than for $\hbar
\omega_0$=$7.90\: $meV ($l_0$=$120\: $\AA) as should be expected since
the electronic density is lower in the latter case and the validity of
the HFA is not so good. Almost the same instability regions are found
in the case of the LDA where correlation effects are neglected
(\ref{LDA}).

\section{Intrasubband Excitations}
\label{s:intrasubband}

When the wave vector component along the wire is not zero,
intrasubband excitations shows up in the spectra.  We will first
consider the case of three subbands occupied.  For the case of two
subbands a comparison with experiments has been briefly communicated
in Ref.\ \onlinecite{Brataas96:325}.  Finally we study how the virtual
transitions to upper nonoccupied subbands may alter the elementary
excitation energies in the Tomonaga-Luttinger model, for only a single
subband occupied.

\subsection{Low- and high-energy excitations: Numerical results}
\label{s:result}

The external parabolic potential was set to $11.37\: $meV
($l_0$=$100\:$\AA) resulting in a self-consistent Hartree-Fock subband
spacing between the lowest subbands of $6.6\: $meV and between the
next lowest subbands of $8.3\: $meV at the zone center.  Three
subbands were occupied with densities $n_0$=9.4$\times10^5\:
$cm$^{-1}$, $n_1$=7.3$\times10^5\: $cm$^{-1}$ and
$n_2$=3.3$\times10^4\: $cm$^{-1}$.  We take the wire to have a length
of $L_x$=$1.0\:\mu$m.

For three subbands occupied it is generally
expected\cite{Li89:5860,Yu90:1496,Li91:11768} that the SDE dispersions are
linear in $q$, with sound velocities $v_{0}$, $v_{1}$ and $v_{2}$ for
intrasubband excitations in the three lowest subbands.  The Fermi
velocities are given by the subband densities $v_{i}$=$\pi \hbar
n_i$/$m^*$ ($i$=0,1,2).  The CDE dispersion also has three
branches. The in-phase mode is approximately\cite{Li89:5860}
\begin{equation}
  \omega_{\rho}^+(q) = |q| \sqrt{2 (v_{0} + v_{1} + v_{2})
    V(q)/\hbar\pi } \, ,
\label{3charge+}
\end{equation}
where $V(q)$ is the Fourier transform of the Coulomb potential
$(e^2/\kappa)[(x$-$x')^2$+$l_0^2]^{-1/2}$.  In addition there are two
out-of-phase modes with linear dispersion.\cite{Li91:11768}

We note in general good qualitative agreement of our numerical results
with the measured Raman spectra.\cite{Goni91:3298,Schmeller94:14778}
At finite wave-vectors along the wire we found low-frequency SDE and
CDE corresponding to one-dimensional intrasubband motion of the
electron gas.  The SDE, as can be seen in Fig.\ \ref{f:intraSDE}
follow a linear dispersion in the long wavelength limit.  For three
subbands occupied we see the three collective spin density modes with
group velocities at $\tilde{v}_0$, $\tilde{v}_1$ and $\tilde{v}_2$.
The Landau damping destroys the higher energy modes at large enough
wave-vectors.  The Landau damping shows up in enhanced intensities of
satellite single-particle peaks around the SDE, as shown in Fig.\
\ref{f:intraSDE}, since we considered a wire of a finite length
($L_x=1.0\:$ $\mu m$).  For infinite wires one should expect a large number
of SPE peaks merging into a broad band.  We found the sound velocity
$\tilde{v}_0$ larger than the corresponding Fermi velocity
($\tilde{v}_0$=1.02$v_0$), and the sound velocities $\tilde{v}_1$ and
$\tilde{v}_2$ smaller than the corresponding Fermi velocities,
($\tilde{v}_1$=$0.98 v_1$ and $\tilde{v}_2$=$0.64 v_2$).  This is
caused by the exchange interaction within each subband and
intersubband coupling giving rise to renormalized sound
velocities.\cite{Schulz93:1864,Brataas96:325} The other factor comes
from intersubband virtual transitions due to exchange interaction
leading to exchange screening of SDE.  These low-energy features
within the HF-RPA must be considered as approximate results.  

The $q$ dependence of the intrasubband plasmon energy (CDE+) can be
fitted by $q[$-$\ln{q}]^{-1/2}$, as shown in the inset in Fig.\
\ref{f:intraCDE}.  That is what is expected for one-dimensional
wires.\cite{Li89:5860,Schulz93:1864} For three subbands occupied there
must also be two CDEs having linear dispersion which may also be seen
in Fig. \ref{f:intraCDE}.  The intensities of these excitations are
however weak and maybe difficult to observe experimentally. The
plasmon decays at higher wave-vectors due to the Landau damping, as
seen in Fig.\ \ref{f:intraCDE}.  

The calculations performed in Refs.\
\onlinecite{Li91:11768,Wendler94:13607,Reboredo94:15174,Hwang94:17267}
within the Hartree random phase approximation have been demonstrated
to agree very well with experimental results for the charge-density
excitations.  In order to see the effect of the vertex correction on
the intrasubband charge-density excitations we have also performed a
calculation with the same sample parameters as above, but omitting the
exchange interaction (not shown).  Indeed, we found that the vertex
corrections have negligible effect on the in-phase intrasubband
plasmon energy.  It agrees within the numerical accuracy (1 \%).  For
the two out-of-phase modes the Hartree calculation gives a Fermi
velocity 15 \% smaller for the highest energy out-of-phase mode and
almost identical for the lowest out-of-phase mode as compared to the
HF-RPA calculation.  The HF-RPA generally gives about twice as large
relative intensity of the out-of-phase modes relative to the in-phase
intensity as compared to the Hartree random phase approximation.

It should be noted that according to our analysis of the wave-vector
dependence of intrasubband SDE the frequency of the lowest energy SDE
decreases at large $q$$\le$$k_2$, and at some value of $q$ close to $2
k_2$ it goes to zero.  That means an intrinsic instability of the
system with respect to formation of spin density waves (SDW) below
some critical temperature.  This is the well-known Peierls instability
of one-dimensional systems.  Overhauser\cite{Overhauser62:1437} showed
that the HF paramagnetic state in 3D systems is always unstable with
respect to formation of a static spin-density wave having a wave
vector $q$$\approx$$2k_F$, and MacDonald\cite{MacDonald85:1003} found
that the same was expected in 2D systems.  Therefore the Hartree-Fock
spin susceptibility of the paramagnetic state has a singularity near
$q$=$2k_F$.  The instability in the quasi 1D wire emerges from our
mean-field analysis, and in the framework of this approach one should
take into account SDW long-range order in the ground state.  However,
exact results making use of the Luttinger model predict no SDW
long-range order, but predict a slowly decaying Wigner crystal of
short range order at $4 k_F$.\cite{Schulz93:1864}

\subsection{Screening in the Tomonaga-Luttinger model}
\label{s:screening}

In the Tomonaga-Luttinger model, the Coulomb interaction is restricted
to processes within the lowest occupied subband and the kinetic energy
is approximated by a linear dispersion.\cite{Haldane81:2585} In this
way the elementary low-energy excitations may be found exactly by
using a bosonization technique.  We will now study how the excitation
energies for the Tomonaga-Luttinger model may change when the coupling
to the higher subbands is taken into account.  The virtual transitions
to the upper subbands will give rise to direct and exchange screening.
A second order perturbation argument shows that contributions from
intersubband transitions will be inversely proportional to the subband
spacing.  We will now give some more quantitative results derived from
an effective low-energy Hamiltonian where the virtual high-energy
intersubband excitations are treated within the HF-RPA.  In quantum
wires used in Raman measurements\cite{Goni91:3298,Schmeller94:14778},
coupling to higher subbands is always present and may not be small.
Since this coupling may renormalize the excitation energies appearing
in the experimental system it is an important physical parameter.  The
renormalization factors may be measured directly in an experiment if
one is able to control the subband spacing (i.e. the curvature of the
parabolic confinement) while keeping the 1D electronic density of the
lowest subband constant.

For the low-energy excitations the interaction may be approximated by
\widetext
\begin{eqnarray}
  \hat{V} & = & \sum_{ij,qkp,s\sigma} [ H^{ij}(k,p)
  \hat{c}_{i\sigma}^{\dag}(p-q/2) \hat{c}_{j \sigma}(p+q/2) \hat{c}_{0
    s}^{\dag}(k+q/2) \hat{c}_{0 s}(k-q/2) \nonumber \\ & - &
  F^{ij}(k,p) \hat{c}_{is}^{\dag}(k+q/2) \hat{c}_{j \sigma}(p+q/2)
  \hat{c}_{0 \sigma}^{\dag}(p-q/2) \hat{c}_{0 s}(k-q/2) ] \, ,
\end{eqnarray}
\narrowtext 
where the first term represents the Hartree interaction
\begin{equation}
  H^{ij}(k,p) = V^{i0;0j}(p,k,k,p)
\end{equation}
and the second the Fock interaction
\begin{equation}
  F^{ij}(k,p) = V^{0i;0j}(k,p,k,p)
\end{equation}
between the lowest subband and the higher subbands.  The matrix
elements are evaluated within the Hartree-Fock basis,
$V^{ij;mn}(kp;qr)$=$\int d{\bbox r}\int d{\bbox r}^{\prime}$
$\psi_{i,k}^*({\bbox r})
\psi_{n,r}({\bbox r})\psi_{j,p}^*({\bbox r}^{\prime})\psi_{m,q}({\bbox r}^{\prime})$/
$|{\bbox r}-{\bbox r}^{\prime}|$.  By adding this interaction part to the TLM
Hamiltonian the virtual transitions to the upper subbands are
included.  Following the standard bosonization
procedure\cite{Haldane81:2585,Mahan90} we now introduce the boson
operators for charge-density excitations $\hat{b}^+$ and spin-density
excitations $\hat{b}^-$ in the lowest subband ($q>0$),
\widetext
\begin{equation}
  \sum_{k>0} \left( \hat{c}_{0\uparrow}^{\dag}(k+q/2)
    \hat{c}_{0\uparrow}(k-q/2) \pm \hat{c}_{0\downarrow}^{\dag}(k+q/2)
    \hat{c}_{0\downarrow}(k-q/2) \right) = \hat{b}^{\pm}(q)
  \sqrt\frac{qL_x}{\pi}
\end{equation}
and 
\begin{equation}
  \sum_{k<0} \left( \hat{c}_{0\uparrow}^{\dag}(k-q/2)
    \hat{c}_{0\uparrow}(k+q/2) \pm \hat{c}_{0\downarrow}^{\dag}(k-q/2)
    \hat{c}_{0\downarrow}(k+q/2) \right) = \hat{b}^{\pm}(-q)
  \sqrt{\frac{qL_x}{\pi}} \, .
\end{equation} 
\narrowtext 
The spin and charge parts of the Hamiltonian decouple so that the
interaction Hamiltonian may be written as
$\hat{V}$=$\hat{V}^{+}$+$\hat{V}^{-}$,
\begin{equation}
  \hat{V}^{\pm} = \sum_{ij,kq} \hat{\psi}_{ij}^{\pm}(k,q) \left(
    X_1^{\pm ij}(k,q) \hat{b}(q) + X_2^{\pm ij}(k,q)
    \hat{b}^{\dag}(-q) \right) \, ,
\label{V+-}
\end{equation}
where we have introduced the intersubband charge-density and
spin-density operators
\begin{eqnarray}
  \hat{\psi}_{ij}^{\pm}(k,q) & = & \hat{c}_{i\uparrow}(k-q/2)
  \hat{c}_{j\uparrow}(k+q/2) \pm \nonumber \\ & &
  \hat{c}_{i\downarrow}(k-q/2) \hat{c}_{j\downarrow}(k+q/2) \, ,
\label{i+-}
\end{eqnarray}
the charge-density matrix element ($n$=1,2)
\begin{equation}
  X_n^{+ij}(k,q) = \sqrt{|\frac{qL_x}{\pi}|} \left( H^{ij}(k_F,k) -
    \frac{1}{2}F_n^{ij}(k,q) \right) \, ,
\label{mat+}
\end{equation}
and the spin-density matrix elements
\begin{equation}
  X_n^{-ij}(k,q) = \sqrt{|\frac{qL_x}{\pi}|} \left( -
    \frac{1}{2}F_n^{ij}(k,q) \right) \, .
\label{mat-}
\end{equation}
Here $F_{n}^{ij}(k,q)$ are combinations of the Fock matrix at $k_F$
and -$k_F$
\begin{equation}
  F_{1}^{ij}(k,q) = F^{ij}(k_F,k) \theta(q) + F^{ij}(-k_F,k)
  \theta(-q)
\end{equation}
and
\begin{equation}
  F_{2}^{ij}(k,q) = F^{ij}(k_F,k) \theta(-q) + F^{ij}(-k_F,k)
  \theta(q) \, .
\end{equation}
There are also spin-flip terms in the interaction Hamiltonian similar
to the spin-spin term in (\ref{V+-}), but the spin triplet is
degenerate so we can limit the discussion to only one of the modes.
The TLM may be solved by performing the Bogoliubov
transformation\cite{Haldane81:2585,Mahan90}
\begin{mathletters}
\begin{equation}
  \hat{b}^{\pm}(q)+\hat{b}^{\pm \dag}(-q) = \frac{\hbar
    \omega(q)}{E^{\pm}(q)} \left( \hat{\beta}^{\pm}(q) +
    \hat{\beta}^{\pm \dag}(-q) \right)
\end{equation}
\begin{equation}
  \hat{b}^{\pm}(q)-\hat{b}^{\pm \dag}(-q) = \frac{\hbar
    \omega(q)}{E^{\pm}(q)} \left( \hat{\beta}^{\pm \dag}(q) -
    \hat{\beta}^{\pm}(-q) \right) \, ,
\end{equation}
\label{bog}
\end{mathletters}
\noindent
where $\omega(q)$=$v_{F}|q|$ is the unperturbed excitation frequency
and $E^{\pm}(q)$ is the excitation energy of the charge-density and
spin-density excitations given by the one-band Luttinger model.  In
order to find the elementary intrasubband excitations in the system we
define the time-ordered boson Green's functions
\begin{equation}
  D^{\pm}(q,\omega) = -\frac{i}{\hbar} \int_{-\infty}^{\infty}
  e^{i\omega t} \langle | T \hat{\beta}^{\pm}(q,t) \hat{\beta}^{\pm
    \dag}(q,0) | \rangle
\label{bosGreen}
\end{equation}
and
\begin{equation}
  D^{\pm}_A(q,\omega) = -\frac{i}{\hbar} \int_{-\infty}^{\infty}
  e^{i\omega t} \langle | T \hat{\beta}^{\pm \dag}(-q,t) \hat{\beta}^{\pm
    \dag}(q,0) | \rangle \, ,
\end{equation}
where the ordinary unperturbed (coupling to the higher subbands is
neglected) Green's functions is $D^{\pm}_0(q,\omega)$=$1/(\hbar
\omega$-$E^{\pm}(q))$ and the anomalous unperturbed Green's function
is $D^{\pm}_{A,0}(q,\omega)$=0.  The interaction Hamiltonian
(\ref{V+-}) may be represented in the same form as (\ref{V+-})
\begin{equation}
  \hat{V}^{\pm} = \sum_q \left[ \hat{f}_1^{\pm 1}(q)
    \hat{\beta}^{\pm}(q) + \hat{f}_2^{\pm 2}(q) \hat{\beta}^{\pm
      \dag}(-q) \right] \, ,
\end{equation}
where $\hat{f}_1^{\pm 1}(q)$ and $\hat{f}_2^{\pm 2}(q)$ are given by
the intersubband operators (\ref{i+-}), the matrix elements
(\ref{mat+}) and (\ref{mat-}), and the transformation (\ref{bog}).  To
lowest order in the intersubband coupling the Dyson's equations are
\begin{eqnarray}
  D^{\pm}(\omega) & = & D^{\pm}_0(\omega) + \nonumber \\ & &
  D^{\pm}_0(\omega) \left[ P_{12}^{\pm} D^{\pm}(\omega) + P_{11}^{\pm}
    D^{\pm}_A(\omega) \right]
\end{eqnarray}
and
\begin{equation}
  D^{\pm}_A(\omega) = 0 + D^{\pm}_0(-\omega) \left[ P_{12}
    D^{\pm}_A(\omega) + P_{11} D^{\pm}(\omega) \right] \, ,
\end{equation}
where the polarizations $P_{\alpha\beta}^{\pm}(q,\omega)$
($\alpha,\beta=1,2$) are defined as
\begin{equation}
  P_{\alpha\beta}^{\pm}(q,\omega) = - \frac{i}{\hbar}
  \int_{-\infty}^{\infty} dt e^{i \omega t} \langle | T
  \hat{f}_{\alpha}^{\pm}(-q,t) \hat{f}_{\beta}^{\pm}(q,0) | \rangle
\label{polarization}
\end{equation}
and may be expressed in terms of the time-ordered two-particle
operator $\Pi^{T\pm}$ given by similar equations as the retarded
two-particle Green's functions $\Pi^{\pm}$ in (\ref{twocharge}).
Intersubband excitations are nearly vertical and the intersubband
excitation energies are much larger than the low-energy intrasubband
excitation energy, hence we approximate the polarizations
(\ref{polarization}) by it's limit when
$(q,\omega)$$\rightarrow$(0,0).
The poles of the Green's function (\ref{bosGreen}) defines the
renormalized intrasubband excitation energies
\begin{equation}
  \tilde{E}^{\pm}(q)^2 = E^{\pm}(q)^2 + 2 P_{12} E^{\pm}(q) +
  (P_{12}^2-P_{11}^2) \, .
\end{equation}
The virtual transitions to the higher subbands are included by the
polarization operator (\ref{polarization}) that describes the coupling
between the lowest occupied subband and all the higher subbands.  By
carrying out this calculation we find that the shift of the excitation
energies may be represented by the dimensionless parameters ($n$=1,2)
\begin{equation}
  h_n^{\pm} = - \frac{2L_x}{v_F \hbar \pi} \sum_{ijlm,kp} W^{\pm
    ij}_n(k) W^{\pm lm}_n(p) \Pi_{ijlm}^{T\pm}(k,p) \, ,
\label{screen+-}
\end{equation}
where $W^{\pm ij}_1(k)$ and $W^{\pm ij}_2(k)$ are given by
\begin{eqnarray}
  W^{+ ij}_{1}(k) & = & H^{ij}(k_F,k) \nonumber \\ & - & \frac{1}{4}
  \left( F^{ij}(k_F,k) + F^{ij}(-k_F,k) \right) \, ,
\end{eqnarray}
\begin{equation}
  W^{- ij}_{1}(k) = - \frac{1}{4} \left( F^{ij}(k_F,k) +
    F^{ij}(-k_F,k) \right) \, ,
\end{equation}
and 
\begin{equation}
  W^{\pm ij}_{2}(k) = - \frac{1}{4} \left( F^{ij}(k_F,k) -
    F^{ij}(-k_F,k) \right) \, .
\label{W2}
\end{equation}
The intrasubband excitation energies are
\begin{equation}
  \left( \frac{\tilde{E}^{\pm}(q)}{E^{\pm}(q)} \right)^2 = \left(1 -
    (\frac{\hbar \omega(q)}{E^{\pm}(q)})^2 \tilde{h}_1^{\pm} \right)
  \left(1 - \tilde{h}_2^{\pm} \right) \, ,
\label{intraren}
\end{equation}
where $h_1^{\pm}$ and $h_2^{\pm}$ represent the contribution caused by
screening due to the virtual excitations to the nonoccupied upper
subbands.  Stability of the system with respect to compression
requires the screening factors not to be too strong.  If not, then the
perturbation approach is not valid since we see from (\ref{intraren})
that too strong screening factors lead to imaginary intrasubband
excitation energies.  For charge-density excitations the plasmon
energy has the low-energy behavior $q[$-$\ln{q}]^{1/2}$ so that the
renormalization factor $h_1^+$ is unimportant.  The screening factors
$h_2^{\pm}$ are given by differences of the Fock matrix elements with
respect to $k_F$ and -$k_F$ (\ref{W2}) and are therefore expected to
be small.

We have evaluated the intersubband screening factors $h_1^{\pm}$ and
$h_2^{\pm}$ for a wire of density
$n_{1D}$=2.0$\times$$10^5\:$cm$^{-1}$ and for subband separation
starting at 2 meV to 46 meV, as shown in Fig.\ (\ref{f:screening}).
The screening factor $h_2^{\pm}$ appears in the calculation orders of
magnitude smaller than $h_1^{\pm}$ and may be neglected, as already
pointed out above.  Even at an external potential separation of
$2\:$meV they are smaller than $1\%$.  Since the effect of the
screening factor $h_1^+$ is unimportant, we find that the plasmon
energy is rather insensitive to the coupling to higher subbands.
Screening has a larger effect on spin-density excitations.  The effect
on the SDE sound velocity is about $5\%$ down to a subband spacing of
4 meV.  For smaller subband spacings, the exchange screening factor
$h_1^-$ appears to decrease, but this is due to the fact that the HFA
is not valid in this small subband separation regime, because the
spin-density excitations are overdamped.  The screening parameters
$h_1^{\pm} $decrease inversely proportional to the subband spacing as
expected, but the factors $h_2^{\pm}$ decrease faster than inversely
proportional to the square of the subband spacing.  The screening
parameters are not very sensitive to the 1D electronic density.  A
similar calculation for $n_{1D}$=3$\times$$10^5\:$cm$^{-1}$ (not
shown) shows that the factor $h_1^+$ decreases by less than 10\% at
$\hbar \omega_0$=$5\:$meV as compared to the results in Fig.
(\ref{f:screening}).  Similarly the factor $h_1^-$ decreases by less
than 20\% at $\hbar \omega_0$=$5\:$meV.

\section{Conclusions}
\label{s:conclusion}

The importance of the vertex corrections (exchange interaction) that
give rise to collective spin-density excitations in quantum wires is
demonstrated. For charge-density excitations vertex corrections are
less important.  The calculated spin-density excitation spectra and
charge-density excitation spectra seem to agree well with experiments.
The dipole intersubband spin-density excitation energy oscillates with
the number of occupied subbands accompanied by peaks in the static
spin-density susceptibility.  For a single subband occupied screening
by virtual transitions to the upper subband renormalizes the
spin-density energy by the same order of magnitude as the correction
due to backscattering mixing right- and left-moving modes.

\acknowledgements{A. B. would like to thank A. Sudb\o\ and L. J. Sham
  for stimulating discussions and M. Willander for a great hospitality
  at the Link\o ping University where parts of this work was carried
  out.  This work has been supported in part by a NorFa grant, the
  Icelandic Science Foundation, the University of Iceland Research
  Fund, the German Science Foundation DFG, and the Research Council of
  Norway (Program for Supercomputing) through a grant of computing
  time.}

\appendix

\section{Singular cut-off}

In an infinite wire the ground state energy of an interacting electron
gas without background charges is divergent, which requires a cut-off
introduced by background charges for the low-wave vector interaction
$q$ similar to the 3D situation.  Consider the Hamilton operator in
the basis of the noninteracting states
\begin{equation}
  \psi_{nk}(x,y) = \frac{1}{\sqrt{L_x}}e^{ikx} \phi_n(y) \, .
\end{equation}
The Coulomb interaction in the noninteracting basis is
\widetext
\begin{equation}
  \hat{H}_{int} = \frac{1}{2} \sum_{imlj} \sum_{kk^{\prime}q}
  \sum_{\sigma \sigma^{\prime}} V^{im;lj}_0(q)
  \hat{c}_{i\sigma}^{\dag}(k-q)
  \hat{c}_{m\sigma^{\prime}}^{\dag}(k^{\prime}+q)
  \hat{c}_{l\sigma^{\prime}}(k^{\prime}) \hat{c}_{j \sigma}(k)
\end{equation}
\narrowtext 
where $\hat{c}_{i\sigma}(k)$ destroys an electron in a state with
transverse quantum number $i$, spin $\sigma$ and longitudinal
wave-vector $k$ and the Coulomb matrix elements are
\begin{eqnarray}
  && V^{im;lj}_0(q) = \nonumber \\ && \frac{e^2}{\kappa} \int d{\bbox r}
  \int d{\bbox r}^{\prime} \frac{1}{L_x^2} \frac{e^{iq(x-x^{\prime})}
    \phi_i(y) \phi_j(y) \phi_m(y^{\prime}) \phi_l(y^{\prime})
    }{|{\bbox r}-{\bbox r}^{\prime}|} \, ,
\end{eqnarray}
which in the limit of long wires can be written as
\begin{eqnarray}
  && V^{im;lj}_0(q) = \frac{2e^2}{\kappa L_x} \times \nonumber \\ &&
  \int dy \int dy^{\prime} K_0(|q(y-y^{\prime})|) \phi_i(y) \phi_j(y)
  \phi_m(y^{\prime}) \phi_l(y^{\prime}) \, .
\label{Vnonint}
\end{eqnarray}
For small arguments the Bessel function is asymptotically
$K_0(x)$$\sim$-$\ln{x}$.  For small wave vectors $q$ we may therefore
take away the divergent part in the matrix element
\begin{eqnarray}
  && V^{im;lj}(q) = - \frac{2e^2}{\kappa L_x} \ln{q l_0} \delta_{i,j}
  \delta_{m,l} - \frac{2e^2}{\kappa L_x} \times \nonumber \\ && \int
  dy \int dy^{\prime} \ln{|\frac{y-y^{\prime}}{l_0}|} \phi_i(y)
  \phi_j(y) \phi_m(y^{\prime}) \phi_l(y^{\prime}) \, .
\end{eqnarray}
For a finite system we may therefore write
\begin{equation}
  V^{im;lj}(q) = \delta_{q,0} \delta_{i,j} \delta_{m,l} C +
  \tilde{V}^{im;lj}(q)
\label{Vlongwire}
\end{equation}
where $C$ is a large constant.  This first term in (\ref{Vlongwire})
gives a part $C(\hat{N}^2$-$\hat{N})$ to the Hamiltonian where
$\hat{N}$ is the number operator of particles.  We consider states
where the total number of particles is a good quantum number so that
this term is cancelled by a similar term from the background charges.
The renormalized matrix element for finite wave vector $q$ is given by
(\ref{Vnonint}) and for $q$=0
\begin{eqnarray}
  && \tilde{V}^{im;lj}(q=0) = - \frac{2e^2}{\kappa L_x} \times
  \nonumber \\ && \int dy \int dy^{\prime}
  \ln{|\frac{y-y^{\prime}}{L}|} \phi_i(y) \phi_j(y) \phi_m(y^{\prime})
  \phi_l(y^{\prime}) \, .
\end{eqnarray}
In the Hartree approximation this is the same as the long wire limit
in Refs.
[\onlinecite{Gudmundsson88:453},\onlinecite{Gudmundsson95:17744}].
For an easier numerical evaluation by using the integral form of the
Bessel function
\begin{equation}
  K_0(x) = \int_{-\infty}^{\infty} dt \frac{e^{ixt}}{2\sqrt{t^2+1}} \,
  ,
\end{equation}
the integral (\ref{Vnonint}) may be written as
\begin{eqnarray}
  && V_0^{im;lj}(q) = \frac{e^2}{\kappa L_x} \times \nonumber \\ &&
  \int_{-\infty}^{\infty} du \frac{1}{\sqrt{u^2 + q^2l_0^2}}
  I_{i,j}(u) I_{m,l}(-u) \, ,
\label{numfock}
\end{eqnarray}
where the function $I_{i,j}(u)$ is the Fourier transform of the
product of a pair of transverse wave-functions
\begin{equation}
  I_{l,l^{\prime}}(u) = \int_{-\infty}^{\infty} dv \phi_{l}(v)
  \phi_{l^{\prime}}(v) e^{i u v t} \, ,
\label{I}
\end{equation}
and $u$ and $v$ are dimensionless variables.  The function (\ref{I})
may be written as a linear combination of Laguerre functions, so only
a one-dimensional integral has to be evaluated numerically in
(\ref{numfock}).

\section{Diagonalization of Excited States}
\label{s:diagonalization}

The eigenequations (\ref{indcharge}) and (\ref{indspin}) are not well
suited for a numerical diagonalization since they are in general not
symmetric.  In the case of the time dependent Hartree approximation
this problem is easily solved due to the local nature of the
approximation.\cite{Brataas96:4797} The nonlocal character of the Fock
term causes some problems.  However the eigenequations may be
transformed to a symmetric form as will be described since this may be
of interest for other calculations based on the Hartree-Fock random
phase approximation.  This appendix therefore covers some of the more
technical aspects of the calculations.

By decoupling (\ref{indcharge}) or (\ref{indspin}) into symmetric and
antisymmetric combinations the eigenvalue equation may be rewritten.
We define the new variables
\begin{mathletters}
\begin{equation}
  y^{S,\pm}_{ab}({\bbox q},\omega) = \frac{K_{ab}^{\pm}({\bbox q},\omega)
    + K_{ba}^{\pm}({\bbox q},\omega)}
  {\sqrt{(\epsilon_a-\epsilon_b)(f_b-f_a)}},
\end{equation}
\begin{equation}
  y^{A,\pm}_{ab}({\bbox q},\omega) = \left(K_{ab}^{\pm}({\bbox q},\omega)
    - K_{ab}^{\pm}({\bbox q},\omega)\right)
  \sqrt{\frac{\epsilon_a-\epsilon_b}{f_b-f_a}}
\end{equation}
\end{mathletters}
and 
\begin{mathletters}
\begin{eqnarray}
  b_{ab}^{S}({\bbox q}) & = & 2 \sqrt{(f_b-fa)(\epsilon_a-\epsilon_b)} [
  \nonumber \\ && \langle a | e^{-i {\bbox q} \cdot {\bbox r}} | b \rangle
  + \langle b | e^{-i {\bbox q} \cdot {\bbox r}} | a \rangle ], \\ 
  b_{ab}^{A}({\bbox q}) & = & 2
  \sqrt{\frac{f_b-fa}{\epsilon_a-\epsilon_b}} \left( \langle a | e^{-i
      {\bbox q} \cdot {\bbox r}} | b \rangle - \langle b | e^{-i {\bbox q}
      \cdot {\bbox r}} | a \rangle \right)
\end{eqnarray}
\end{mathletters}
\noindent
and restrict the summation to $\epsilon_a < \epsilon_b$ obtaining the
matrix equations
\begin{mathletters}
\label{as}
\begin{equation}
  - \omega {\bbox y}^{A,\pm}({\bbox q},\omega) = {\cal S}^{\pm} \cdot
  {\bbox y}^{S,\pm}({\bbox q},\omega) + {\bbox b}^{S,\pm}({\bbox q})
\end{equation}
\begin{equation}
- \omega {\bbox y}^{S,\pm}({\bbox q},\omega) = {\cal A}^{\pm} 
\cdot {\bbox y}^{A,\pm}({\bbox q},\omega) + {\bbox b}^{A,\pm}({\bbox q}) \, ,
\end{equation}
\end{mathletters}
\noindent
where the matrices (since all Coulomb matrix elements are real) ${\cal
  S}^{\pm}$ and ${\cal A}^{\pm}$ are symmetric and antisymmetric
combinations of the Coulomb interaction elements defined by
\begin{eqnarray}
  {\cal S}^{+}_{ab,cd} & = &
  \sqrt{(f_b-f_a)(f_d-f_c)(\epsilon_a-\epsilon_b)(\epsilon_c-\epsilon_d)}
  [ \nonumber \\ & & 2(V_{cb;ad} + V_{ca;bd} + V_{db;ac} + V_{da;bc})
  \nonumber \\ &- & (V_{bc;ad} + V_{ac;bd} + V_{bd;ac} + V_{ad;bc})]
\end{eqnarray}
and
\begin{eqnarray}
  {\cal A}^{+}_{ab,cd} & = & \sqrt{\frac{(f_b-f_a)(f_d-f_c)}
    {(\epsilon_a-\epsilon_b)(\epsilon_c-\epsilon_d)}}[ \nonumber \\ &
  & 2(V_{cb;ad} - V_{ca;bd} - V_{db;ac} + V_{da;bc}) \nonumber \\ & -
  & (V_{bc;ad} - V_{ac;bd} - V_{bd;ac} + V_{ad;bc})]
\end{eqnarray}
Similarly for the spin-density excitations the Hartree terms are
absent.  By now assuming that the matrix ${\cal A}$ is positive
definite (this assumption has always been satisfied in our
calculations), we can use a Cholesky decomposition ${\cal A}$=${\cal
  L}$$\cdot$${\cal L}^T$ where ${\cal L}$ is a real matrix and
diagonalize the system.  The excitation energies are given by the
symmetric eigenvalue problem
\begin{equation}
  {\cal L}^T \cdot {\cal S} \cdot {\cal L} \cdot {\cal Z} = {\cal Z}
  \cdot \omega^2_* \, ,
\end{equation}
where the matrix ${\cal Z}$ contains the eigenvectors of the
corresponding eigenvalues in the diagonal matrix $\omega_*^2$.  If at
least one eigenvalue in the matrix $\omega_*^2$ is negative, it means
that we have overdamped modes.  The Hartree-Fock ground state is then
unstable.

The imaginary part of the charge-density and spin-density correlation
functions can now be found to be
\begin{eqnarray}
  S({\bbox q},\omega) & \equiv & - \text{Im} \left( \chi({\bbox q},\omega)
  \right) \nonumber \\ & = & \pm \frac{\pi}{4} \sum_x \delta(\omega
  \mp \omega_x^*) \left(W_x^S({\bbox q}) \mp W_x^A({\bbox q}) \right)^2 \,
  ,
\label{intensity}
\end{eqnarray}
where the weights $W_x^S({\bbox q})$ and $W_x^S({\bbox q})$ are defined as
\begin{equation}
  W_x^A({\bbox q}) = \sum_{yz} b_y^A({\bbox q}) \left(L_{yz}^T\right)^{-1}
  {\cal Z}_{zx} \sqrt{\omega_x^*}
\end{equation}
\begin{equation}
  W_x^S({\bbox q}) = \sum_{yz} b_y^S({\bbox q}) L_{yz} {\cal Z}_{zx}
  \sqrt{1/\omega_x^*} \, ,
\end{equation}
and the indices $x,y,z$ denotes electron-hole pair excitations.  From
(\ref{intensity}) it is seen that the charge-density and spin-density
excitation intensities are positive as they should be.  In general,
the spectral function should have the symmetry
$S({\bbox q},\omega)$=-$S($-${\bbox q},$-$\omega)$ since from
(\ref{Ccorr}) and (\ref{Scorr}) we see that
$\chi({\bbox q},\omega)$=$\chi^*($-${\bbox q},$-$\omega)$.  Here the
weights have the symmetry $W_x^S({\bbox q})$=$W_x^S($-${\bbox q})$ and
$W_x^A({\bbox q})$=-$W_x^A($-${\bbox q})$ so that (\ref{intensity}) has
the correct symmetry.

The real part of the correlation function 
\begin{equation}
  R({\bbox q},\omega) = -\text{Re} \left( \chi({\bbox q},\omega) \right)
\end{equation}
may be found from the imaginary part by the Kramers-Kronig relation
\begin{equation}
  R({\bbox q},\omega) = \frac{1}{\pi} \int_{-\infty}^{\infty}
  d\omega^{\prime} S({\bbox q},\omega^{\prime}) \mbox{P}
  \frac{1}{\omega^{\prime}-\omega} \, ,
\end{equation}
where $\mbox{P}$ denotes the principal part.  In the static case the
susceptibility is therefore
\begin{equation}
  R({\bbox q}) = \frac{1}{2} \sum_x \frac{(W_x^S({\bbox q}))^2 +
    (W_x^A({\bbox q}))^2} {\omega_x^*} \, .
\end{equation}
The static susceptibility diverges if there are excitations with very
low excitation energy and finite weight.

\begin{figure}
\caption{
  Feynman diagrams for the single-particle Green's function in the
  Hartree-Fock Approximation.  The thick lines represent the
  Hartree-Fock single-particle propagator, the thin lines represent
  the noninteracting single-particle propagator and the dashed line is
  the electron-electron interaction.  }
\label{f:HFAground}
\end{figure}

\begin{figure}
\caption{
  Feynman diagrams for the two-particle Green's function in the
  Hartree-Fock random phase approximation.  The shaded boxes represent
  the four point vertex function, the thick lines represent the
  Hartree-Fock single-particle Green's function, and the dashed line
  is the electron-electron interaction.  }
\label{f:HFRPA}
\end{figure}

\begin{figure}
\caption{
  Intersubband spin-density excitations in the dipole approximation as
  a function of the electronic density.  The upper panel shows the
  static susceptibility, the center graph the excitation energy of the
  dominant peak in the spectra, and the lowest subfigure the
  difference between the single-particle energy at the zone center and
  the chemical potential.  The confinement energy is $\hbar
  \omega_0$=11.37 meV and the length of the wire $L_x$=2400 \AA,
  $T$=1.0K, $m^*$=0.067$m_0$ and $\kappa$=12.4.}
\label{f:interSDE11.37}
\end{figure}

\begin{figure}
\caption{
  The same as in Fig.\ (\ref{f:interSDE11.37}), except that the confinement energy is $\hbar \omega_0$ = 7.90 meV.}
\label{f:interSDE7.90}
\end{figure}

\begin{figure}
\caption{
  The energy of the strongest intersubband spin-density excitation in
  the dipole approximation as a function of the electron density for
  $\hbar\omega_0$=7.90 meV (upper panel), and $\hbar\omega_0$=11.37
  meV (lower panel). {\it Only here}, the local density approximation
  has been used for the exchange interaction. Other parameters are as
  in Fig.\ \protect{\ref{f:interSDE7.90}}.}
\label{f:interSDE_CS}
\end{figure}

\begin{figure}
\caption{
  Intrasubband spin-density excitations.  Three subbands were occupied
  in the wire of length $L_x$=$1.0\: \mu$m with densities
  $n_0$=9.4$\times$10$^5\: $cm$^{-1}$, $n_1$=7.3$\times$10$^5\:
  $cm$^{-1}$ and $n_2$=3.3$\times10^5\: $cm$^{-1}$.  The external
  parabolic potential was $\hbar \omega_0$=$11.37\: $meV ($l_0$=$100\:
  $\AA ).  The inset shows the dispersion of the collective
  excitations.  }
\label{f:intraSDE}
\end{figure}

\begin{figure}
\caption{
  Intrasubband charge-density excitations.  Three subbands were
  occupied in the wire of length $L_x$=$1.0\: \mu$m with densities
  $n_0$=9.4$\times$10$^5\: $cm$^{-1}$, $n_1$=7.3$\times$10$^5\:
  $cm$^{-1}$ and $n_2$=3.3$\times10^5\: $cm$^{-1}$.  The external
  parabolic potential was $\hbar \omega_0$=$11.37\: $meV ($l_0$=$100\:
  $\AA ).  The inset shows the dispersion of the collective
  excitations.  }
\label{f:intraCDE}
\end{figure}

\begin{figure}
\caption{
  Screening parameters ($h_1^+$, $h_2^+$, $h_1^-$ and $h_2^-$) as a
  function of the external parabolic confinement energy $\hbar \omega_0$.
  The dashed lines show the charge-density screening parameters
  $h_1^+$ and $h_2^+$ where $h_1^+$ is larger than $h_2^+$.  The
  straight lines show the spin-density screening parameters $h_1^-$
  and $h_2^-$ where $h_1^-$ is larger than $h_2^-$.  The electronic
  density of the wire is $n$=2.0$\times$10$^5$ cm$^{-1}$.
  }
\label{f:screening}
\end{figure}


\begin{thebibliography}{10}

\bibitem{Egeler90:1804}
T. Egeler {\it et~al.}, Phys. Rev. Lett. {\bf 65},  1804  (1990).

\bibitem{Goni91:3298}
A.~R. Go{\~n}i {\it et~al.}, Phys. Rev. Lett. {\bf 67},  3298  (1991).

\bibitem{Schmeller94:14778}
A. Schmeller {\it et~al.}, Phys. Rev. B {\bf 49},  14778  (1994).

\bibitem{Li89:5860}
Q. Li and S.~D. Sarma, Phys. Rev. B {\bf 40},  5860  (1989).

\bibitem{Yu90:1496}
H. Yu and J.~ C. Hermanson, Phys. Rev. B {\bf 42}, 1496 (1990).

\bibitem{Li91:11768}
Q. Li and S.~D. Sarma, Phys. Rev. B {\bf 43}, 11768 (1991).

\bibitem{Wendler94:13607}
L. Wendler and V.~G. Grigoryan, Phys. Rev. B {\bf 49},  13607  (1994).

\bibitem{Reboredo94:15174}
F.~A. Reboredo and C.~R. Proetto, Phys. Rev. B {\bf 50},  15174  (1994).

\bibitem{Hwang94:17267}
E. Hwang and S.~D. Sarma, Phys. Rev. B {\bf 50},  17267  (1994).

\bibitem{Ando82:3893}
T. Ando, J. Phys. Soc. Jpn. {\bf 51},  3893  (1982).

\bibitem{Katayama84:1615}
S. Katayama and T. Ando, J. Phys. Soc. Jpn {\bf 54},  1615  (1984).

\bibitem{Tselis84:3318}
A.~C. Tselis and J.~J. Quinn, Phys. Rev. B {\bf 29},  3318  (1984).

\bibitem{Eliasson87:5569}
G. Eliasson, P. Hawrylak, and J.~J. Quinn, Phys. Rev. B {\bf 35},  5569
  (1987).

\bibitem{Kohn61:1242}
W. Kohn, Phys. Rev. {\bf 123},  1242  (1961).

\bibitem{Mattis65:304}
D.~C. Mattis and E.~H. Lieb, J. Math. Phys. {\bf 6},  304  (1965).

\bibitem{Dzyaloshinskii73:411}
I.~E. Dzyaloshinskii and A.~I. Larkin, Zh. Eksp. Teor. Fiz {\bf 65},  411
  (1973), [Sov. Phys. JETP {\bf 38}, 202 (1974)].

\bibitem{Li92:13713}
Q.~P. Li, S.~D. Sarma, and R. Joynt, Phys. Rev. B {\bf 45},  13713  (1992).

\bibitem{Tomonaga50:544}
S. Tomonaga, Prog. Theor. Phys. (Kyoto) {\bf 5},  544  (1950).

\bibitem{Luttinger63:1154}
J.~M. Luttinger, J. Math. Phys. {\bf 4},  1154  (1963).

\bibitem{Haldane81:2585}
F.~D.~M. Haldane, J. Phys. C {\bf 14},  2585  (1981).

\bibitem{Luther74:589}
A. Luther and V.~J. Emery, Phys. Rev. Lett. {\bf 33},  589  (1974).

\bibitem{Schulz93:1864}
H.~J. Schulz, Phys. Rev. Lett. {\bf 71},  1864  (1993).

\bibitem{Hamilton69}
D.~C. Hamilton and A.~L. McWhorter,  in {\em Light Scattering Spectra of
  Solids}, edited by G.~B. Wright (Springer, New York, 1969).

\bibitem{Kallin84:5655}
C. Kallin and B.~I. Halperin, Phys. Rev. B {\bf 30},  5655  (1984).

\bibitem{MacDonald85:1003}
A.~H. MacDonald, J. Phys. C: Solid State Phys. {\bf 18},  1003  (1985).

\bibitem{Gudmundsson95:11266}
V. Gudmundsson and J.~J. Palacios, Phys. Rev. B {\bf 52},  11266  (1995).

\bibitem{Baym61:287}
G. Baym and L.~P. Kadanoff, Phys. Rev. {\bf 124},  287  (1961).

\bibitem{Mahan90}
G.~D. Mahan, {\em Many-Particle Physics} (Plenum Press, New York, 1990).

\bibitem{Tanatar89:5005}
B. Tanatar and D.~M. Ceperley, Phys. Rev. B {\bf 39},  5005  (1989).

\bibitem{Brataas96:325}
A. Brataas, A.~G. Mal'shukov, V. Gudmundsson, and K.~A. Chao, J. Phys.:
  Condens. Matter {\bf 8},  L325  (1996).

\bibitem{Overhauser62:1437}
A.~W. Overhauser, Phys. Rev. {\bf 128},  1437  (1962).

\bibitem{Hu93:5469}
B.~Y.-K. Hu and S.~D. Sarma, Phys. Rev. B {\bf 48},  5469  (1993).

\bibitem{Gudmundsson88:453}
V. Gudmundsson, R.~R. Gerhardts, R. Johnston, and L. Schweitzer, Z. Phys. B
  {\bf 70},  453  (1988).

\bibitem{Gudmundsson95:17744}
V. Gudmundsson {\it et~al.}, Phys. Rev. B {\bf 51},  17744  (1995).

\bibitem{Brataas96:4797}
A. Brataas, V. Gudmundsson, A.~G. Mal'shukov, and K.~A. Chao, J. Phys.:
  Condens. Matter {\bf 8},  4797  (1996).

\end{thebibliography}
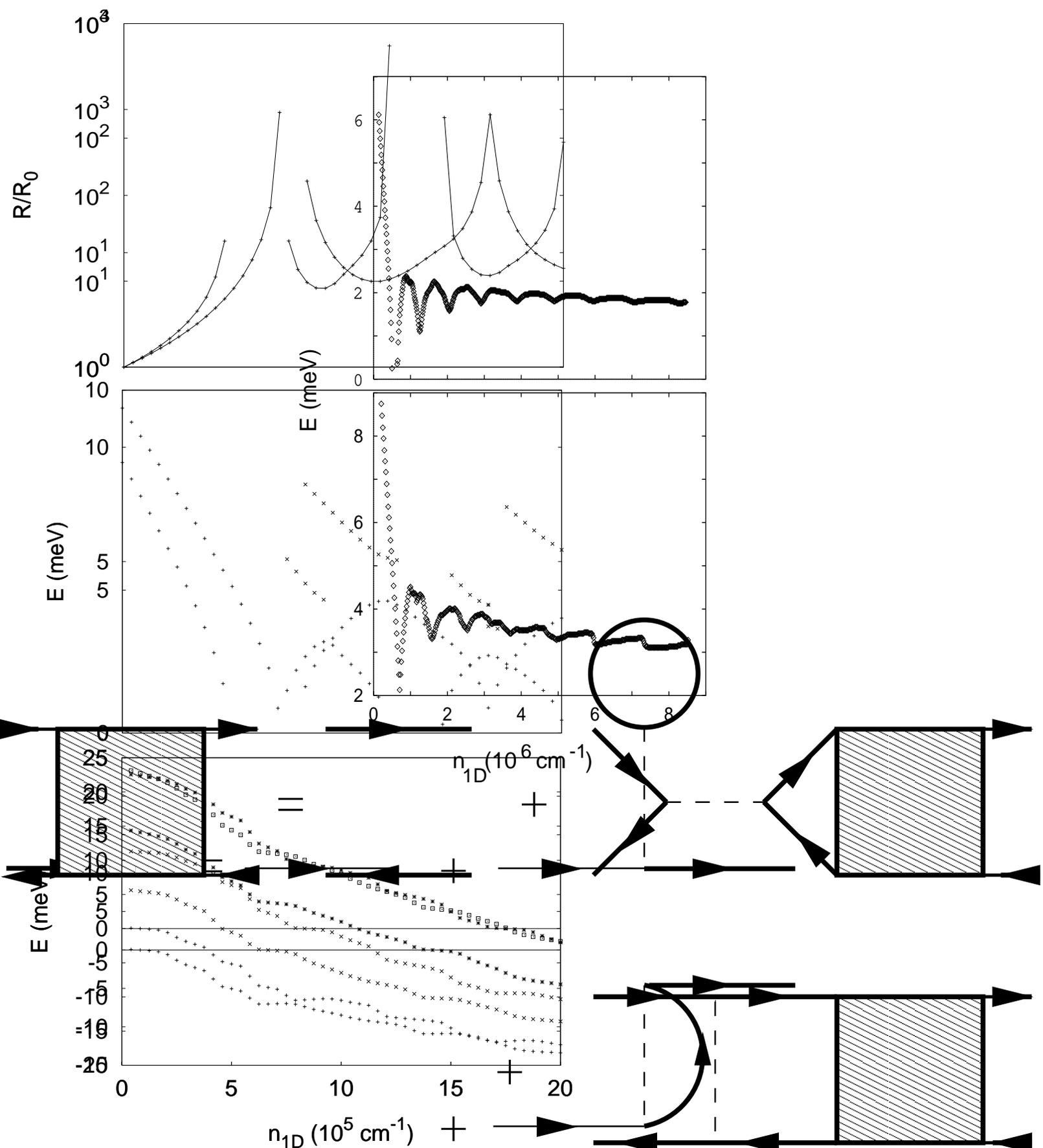
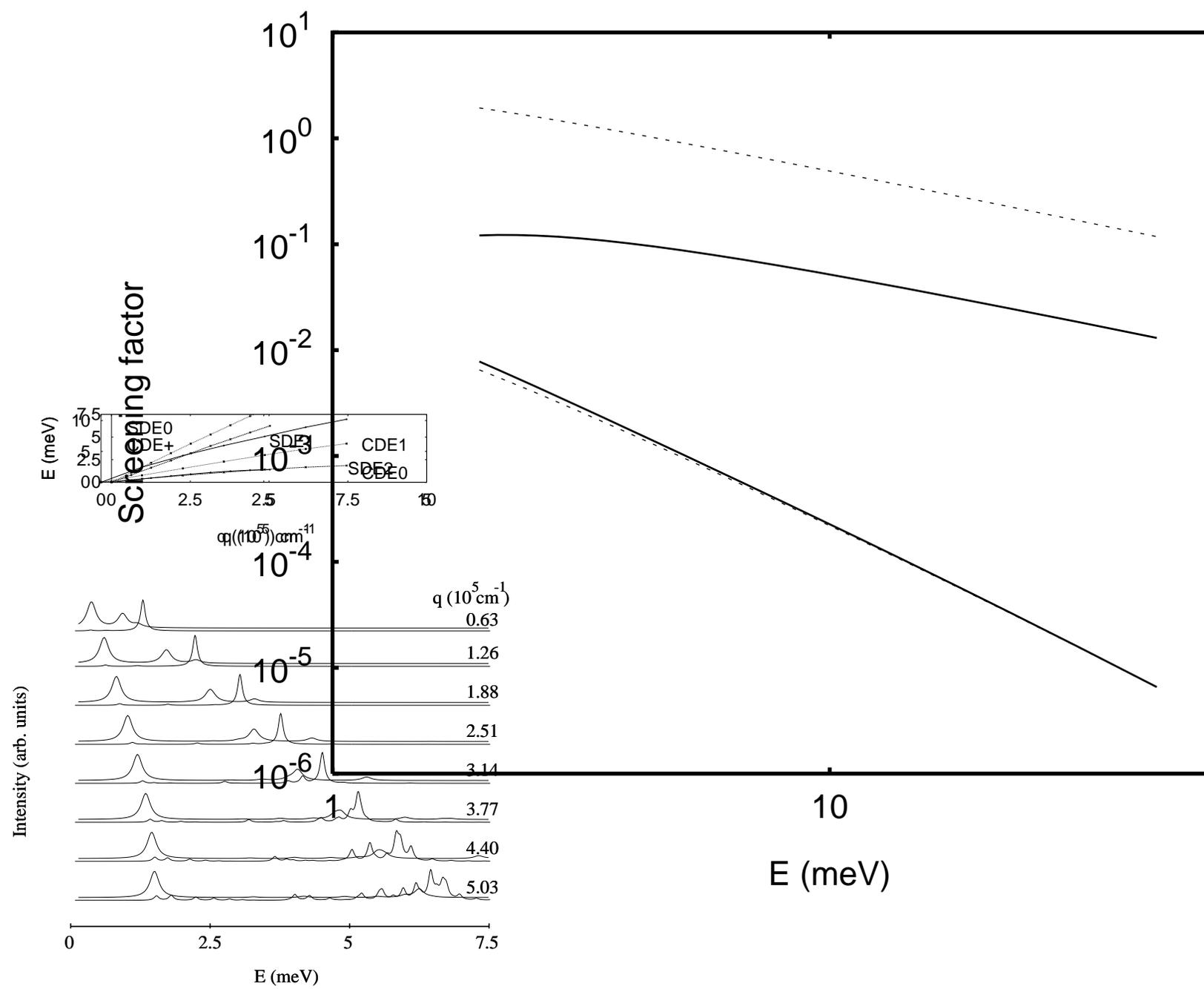
\end{document}